# Dual-band metacomposites containing hybrid Fe and Co-based ferromagnetic microwires

Y. Luo[1], H.X. Peng[1,a)], F.X. Qin[2,b)], M. Ipatov[3], V. Zhukova[3], A. Zhukov[3], J. Gonzalez[3]

[1]*Advanced Composite Centre for Innovation and Science, Department of Aerospace Engineering, University of Bristol, University Walk, Bristol, BS8 1TR, UK*
[2]*1D Nanomaterials Group, National Institute for Material Science, 1-2-1 Sengen, Tsukuba, Ibaraki 305-0047, Japan*
[3]*Dpto. de Fisica de Materiales, Fac. Quimicas, Universidad del Pais Vasco, San Sebastian, 20009, Spain*

We investigated the microwave properties of polymer based metacomposites containing hybridized parallel Fe- and Co-based microwire arrays. A dual-band left-handed feature was observed in the frequency bands of 1.5 to 5.5 GHz and 9 to 17 GHz, indicated by two transmission windows associated with ferromagnetic resonance of Fe-based microwires and long range dipolar resonance between the wire arrays. The plasma frequency after hybridization is significantly increased due to the enhanced effective diameter through the wire-wire interactions between the Fe- and Co-microwire couples. These results offer essential perspectives in designing the multi-band metamaterial for microwave applications such as sensors and cloaking devices.

a) Corresponding author: h.x.peng@bristol.ac.uk(HXP)
b) Corresponding author：faxiang.qin@gmail.com(FXQ)

Left-handed metamaterials possessing exotic electromagnetic (EM) properties that are unavailable in nature have prompted considerable research interest since the initial experimental realization using period metallic wires and split ring resonators (SRRs) by Smith et al.[1] Hitherto, many other resonators or structures such as S-shaped resonators,[2] staple-shaped resonators,[3] paired nanorods[4] and fishnet structures[5] have been deployed to study the conventional metamaterial characteristics in certain frequency range. One critical factor restricting conventional metamaterials in practical applications is the narrow double negative (DNG) frequency band. This necessarily extends the current research to the investigations on broadening the operating frequency band without jeopardizing the metamaterial characteristics. Most recently, dual-band,[6,7,8] multi-band[9,10] and band-tunable metamaterials[11] have been reported to enhance the DNG operating frequencies. Nevertheless, these reported metamaterials with broadened frequency bands lacks sufficient applicability because they were obtained either via the structural destructive interference at the sacrifice of the intensively increased complexity of the material structure, or by employing expensive components in the nano-scale structure to realize the EM transparency in THz frequencies.

To date, it has been experimentally confirmed that the free standing Co-based microwire arrays can excite magnetic field-tunable metamaterial features in a waveguide.[12,13] However, this metamaterial is in fact a meta-structure rather than a true material. On the other hand, polymer composites containing ferromagnetic microwires have demonstrated merits such as low functional filler concentration, high mechanical stress resistivity and outstanding sensing properties.[14] Besides, we have introduced the term of 'metacomposite' to define the microwire-polymer metamaterial and reported upon the realization of single band DNG feature by employing parallel Fe-based microwire arrays in such composites.[15] As Co-based and Fe-based wires have distinctly different magnetic and microwave properties, incorporating both Co-based and Fe-based microwire arrays into polymer matrix can be a clever strategy to broaden the metamaterial operating frequency range, in that different magnetic resonance patterns can be achieved and the dielectric response can be manipulated by managing the mesostructure. In this Letter, we study the microwave behavior of polymer metacomposites containing two hybrid layers of parallel Fe-based and Co-based microwire arrays with/without the external dc magnetic fields up to 3kOe. The dual-band DNG characteristics have been revealed in the hybrid microwire metacomposite and other key findings are: (i) the ferromagnetic resonance (FMR) of Fe-based microwires is found to be responsible for the observed transmission window in the frequency band of 1.5 to 5.5 GHz; (ii) the long-range dipolar resonance between Fe-based and Co-based microwire arrays is identified to be the root cause of the induced transparency at the higher frequency band of 9 ~ 17 GHz; (iii) the increased effective diameter is proposed to account for the observed significantly enhanced plasma frequency.

Experimentally, amorphous glass-coated ferromagnetic microwires $Fe_{77}Si_{10}B_{10}C_3$ (total diameter of 20 μm, glass coat thickness of 3.4 μm) and $Fe_{3.85}Co_{67.05}Ni_{1.44}B_{11.53}Si_{14.47}Mo_{1.66}$ (total diameter of 19.4 μm, glass coat thickness of 3.43 μm) were fabricated by Taylor-Ulitovskiy technique and supplied by TAMAG, Spain.[16,17] To fabricate the hybridized microwire metacomposite, 950 E-glass epoxy prepregs were used as the base material; the Fe-based and Co-based microwires were embedded in a parallel fashion oriented along the glass fibre direction in two respective neighboring prepregs with the wire spacing of 10 mm, two additional layers of prepreg were added on the top and bottom (Fig. 1(a)). To avoid large reflection loss from the wire array superposition, Fe-wire array and Co-wire array were purposely offset by approximate 1 mm (Fig. 1(b)). This was followed by a standard autoclave curing protocol to produce a sample with an in-plane size of 500×500mm$^2$ and 1.5 mm thick. The relevant information regarding the curing procedure can be found in Ref[18]. For comparison, composites containing single microwire layer were also manufactured by planting parallel Fe-based or Co-based wires alone into prepregs with the fixed 10 mm spacing before running the same curing protocol. The microwave properties of the composites were examined by the free space measurement with electrical component $E_k$ along the microwire direction. The related technical details can be found elsewhere.[19] The *S*-parameters were measured in the frequency band of 0.9 to 17 GHz and an external magnetic bias up to 3000 Oe was applied along the microwire direction to investigate the field effect on the microwave behavior. An in-built Reflection/Transmission Epsilon Fast Model was employed to compute the complex permittivity of our major interest.

We exhibit the EM parameters of the microwire composites in the frequency band of 0.9 to 7 GHz in Fig. 2. First, the Co-based wire composite displays the weak transmission and field-tunable properties, which is consistent with the overall high reflection loss shown in Fig. 2(b). This suggests that single Co-based microwire array with a spacing of 10 mm acts as a stop-band component to the incident EM waves in the composite. Further, the transmission and absorption of Fe-based microwire composite are significantly larger (Fig. 2(a) and 2(c)). Noting that the natural ferromagnetic resonance (NFMR) of the used Fe-based wires is realized at 2.3 GHz.[15] As the intrinsic impedance calculates as $Z = (\mu/\varepsilon)^{1/2}$, where $\mu$ and $\varepsilon$ denote respectively the permeability and permittivity, the impedance match of the composite containing Fe-based microwire array is thus significantly improved due to the additional permeability contribution from the microwire NFMR, hence the transmission and absorption. Strikingly, a transmission window is identified with the Co-based wire array added to the Fe-based microwire composite in the frequency band of 1.5 to 5.5 GHz in the absence of dc field, experimentally proving the DNG characteristic of composite containing the hybrid arrays of Fe- and Co-based microwires that has not been reported before. It should be reminded that this behavior resembles that of Fe-based microwire composite with the natural DNG feature,[15] also demonstrating the advantage that extra external stimuli is not required to excite such effect. To explain this left-handed behavior, it was established that the plasma frequency ($f_p$) and FMR frequency physically determine the origin of metamaterial and consequently their bandwidth. By introducing the additional Co-based wire array into the composite, the $f_p$ is enhanced due to the increase of the wire medium complexity therein,[20] as also evidenced in Fig. 4, resulting in the negative permittivity dispersion in the lower frequency band. Meanwhile, the negative permeability should be maintained in the frequencies between 2.3 GHz (NFMR) and ferromagnetic anti-resonance frequencies (FMAR) (higher than the whole measuring frequencies) of the Fe-based wires. In particular, a slightly different window position where the metacomposite features emerge is noted, i.e., the transmission window initiates from 1.5 GHz rather than 2.3 GHz. This small deviation can be attributed to the frozen-in stress during microwires fabrication[21,22] and the effect from polymer matrix via interface after curing that is likely to influence the magnetic properties. As is well acknowledged, Co-based microwires have a much smaller NFMR frequency compared to Fe-based ones but the resonance peak position can be effectively tuned by the external magnetic bias.[23,17] However, the invariant transmission window position with respect to external fields implies that the left-hand behavior is independent on the intrinsic properties of Co-based microwires at low frequency regime (Fig. 2(a)). This is due to the mitigated dynamic coupling between the EM waves and the single Co-based microwire array arising from the large dielectric loss as seen in Fig. 2(b). One might expect to extend the transmission window to even higher frequencies by decreasing the spacing of adjacent Fe-based wires, in the hope of increasing $f_p$ and receiving a stronger magnetic excitation from the composite with higher wire content. However, further investigations on Fe-based wire arrays with spacing of 7 mm and 3 mm, as well as 10 mm Co-based wire array reveal that the transmission window is cancelled out (*results not shown here*). This is because that both Co- and Fe-based wires show such large reflection level that prohibits the transmission window.

Figure 3 displays the transmission, reflection and absorption coefficients of the microwire composites in the frequency band of 7 to 17 GHz. Overall, the strong transmission and reflection of Co-based wire array composite persists, owing to the drastically dwarfed skin depth at higher frequencies and consequently large magnetic and dielectric losses.[24] A remarkable feature is that the transmission window in addition to the reflection and absorption dips have been acquired in the frequencies of 9 to 17 GHz for the Fe-/Co-based hybrid wire array composite, indicating a natural DNG feature at such high frequencies. An absorption peak at 8.5 GHz is also noted, which implies that the magnetic resonance occurred and is responsible for the identified high frequency EM wave-induced transparency. We have elucidated in our earlier study that decreasing wire spacing to 3 mm would induce strong dynamic wire-wire interactions with microwave and realize the long range dipolar resonance.[15] Hence, further reducing the wire spacing to 1 mm, i.e., the spacing between Fe-based and Co-based microwire arrays in the present study would also generate such effect and arouse the noted interaction absorption. Moreover, the circumferential fields created by the coupling between the $E_k$ and the longitudinal anisotropic field of Fe-based microwire array in favor of FMR[25] can also interact with the circumferential anisotropic field of Co-based microwire array, thus enhancing magnetic excitation to some extent. At this point, this interactive magnetic resonance indicates a negative permeability dispersion above the resonance frequency. The observed transmission window also suggests a negative permittivity linking to the $f_p$ that will be discussed later. In addition, it shows that for the single Fe- or Co-based microwire array composite, there is no such effect; this is because that the 10 mm spacing between microwires is too wide to realize meaningful dynamic wire-wire interactions hence the magnetic resonance. Overall, this high-frequency transparency achieved by proper wire misalignment/offset and the dynamic excitation from propagating EM waves broadens the metacomposite operating frequencies and provides essential guides for the metamaterial design. In another perspective, according to Kittel's relations,[26] the above two noted magnetic resonance peaks and transmission windows should have blueshifted with increasing external field. However, the experimental results support that they seem to be independent of dc fields. This can be explained by the degraded magnetic performance of the microwires, conspiring to the formation of hard phases on the surface after experiencing the curing cycle at the high temperature and pressure.[27,28] This issue could be addressed through the proper pretreatment to microwires such as magnetic field[29] or current annealing[30] to anticipate better static and dynamic EM properties.

To understand the dielectric behavior of the identified dual-band metacomposite characteristics, Fig. 4 shows the effective permittivity of microwire composites in the presence of external field up to 3k Oe. The dielectric response of the continuous microwire composite system can be described as the dilute plasmonic behavior and interpreted via $f_p = \sqrt{\frac{c^2}{2\pi b^2 ln\left(\frac{b}{a}\right)}}$, where $c=3\times10^8$ m/s, $b$ and $a$ denote the vacuum light velocity, wire spacing and diameter, respectively. Substituting $b=$ 10mm, $a_{Co}$=15.97 μm and $a_{Fe}$=16.6 μm (inner core diameter), we have the $f_{p,Co}$=4.9 GHz and $f_{p,Fe}$=4.8 GHz, respectively. Obviously, compared with the measured value for the single Fe- or Co-based microwire array composite (Fig. 4), the calculated $f_p$ is overestimated by the equation. We have resolved this issue by introducing an effective diameter, $a_{eff}$, to account for the volume of the circumferential domain that can interplay with the $E_k$ of the microwave, i.e., $f_p^2 = \frac{c^2}{2\pi b^2 ln\left(\frac{b}{a_{eff}}\right)}$.[15] In this sense, the $f_p$ of Co-based wire array composite is slightly higher than that of Fe-based ones (but still much lower than the theoretical value) because they have larger circumferential domain volume, as confirmed in Fig. 4. This compromised $f_p$ fundamentally clarifies the issue why the left-handed features are not accessible in the single Fe- or Co-based microwire array composites. Clearly, a prominent increase of the $f_p$ is obtained with the hybridisation of Fe- and Co-based microwire arrays in the composite, i.e., a negative permittivity dispersion in the whole measuring frequencies (Fig. 4(a)). As per the above modified equation, the effective diameter is closely associated with the intrinsic domain structure of microwires. In the hybrid microwire metacomposite system, taking into account the small spacing (~ one prepreg thick) between the Co- and Fe-based wire layers, the two neighbouring Fe- and Co-based wires can be regarded as a wire pair/couple that interacts with the microwave (Fig. 1), that is to say, the hybridized composite effectively consists of 50 wire couples with approximately 10 mm spacing. As such, the long range dipolar resonance arising from dynamic wire interactions can greatly enhance the dielectric excitation of the wire pair unit at higher frequencies, thus improving the $a_{eff}$ and consequently the $f_p$. Further investigation is still ongoing to address the quantitative relation between wire pair spacing and $f_p$ enhancement. In polymer metacomposites, engineering the $f_p$ towards higher frequencies is always a critical task in the frontier of metamaterial designing. To reduce the wire spacing or to elevate the metamaterial building block's geometrical dimensions appears to be feasible solutions, but neither of these would effectively suppress the excessive reflection loss or the complexity in the fabrication process due to the massive amount of necessary functional units. This negative aspect even makes the established dilute medium model inapplicable.[14,20] In this context, the hybridisation of different microwire arrays in the microwire-composite system, as presented in the present work, paves a new path to enhance the DNG frequency band, which offers more degrees of freedom in the metacomposite design and manufacturing.

In conclusion, we have demonstrated that a new family of metacomposites comprising of hybrid parallel Fe- and Co-based microwire arrays exhibits dual-band left-handed features evidenced by the observed two transmission windows. The plasma frequency has been significantly elevated after the hybridisation of the two arrays in the composites due to the compensation of the effective diameter arising from the dynamic wire-wire interactions. These results demonstrate that this type of hybrid microwires metacomposites is promising for sensing and cloaking applications.

**Acknowledgements**

Yang Luo would like to acknowledge the financial support from University of Bristol Postgraduate Scholarship and China Scholarship Council. FXQ is supported under the JSPS fellowship and Grants-in-Aid for Scientific Research No. 25-03205.

**Figure captions:**

Fig. 1 (Colour online) Schematic illustration of (a) the manufacturing process of metacomposite containing hybrid parallel Fe-based and Co-based microwire arrays with wire spacing of 10 mm and (b) offset between the Fe-wire and Co-wire layer of about 1mm.

Fig. 2 (Colour online) (a) Transmission, (b) reflection and (c) absorption coefficients of the microwire composites containing different microwire arrays in the frequency band of 0.9 to 7 GHz.

Fig. 3 (Colour online) (a) Transmission, (b) reflection and (c) absorption coefficients of the microwire composites containing different microwire arrays in the frequency band of 7 to 16 GHz.

Fig. 4 (Colour online) Frequency plots of (a) real part $\varepsilon'$ and (b) imaginary part $\varepsilon''$ of effective permittivity of the different microwire array composites.

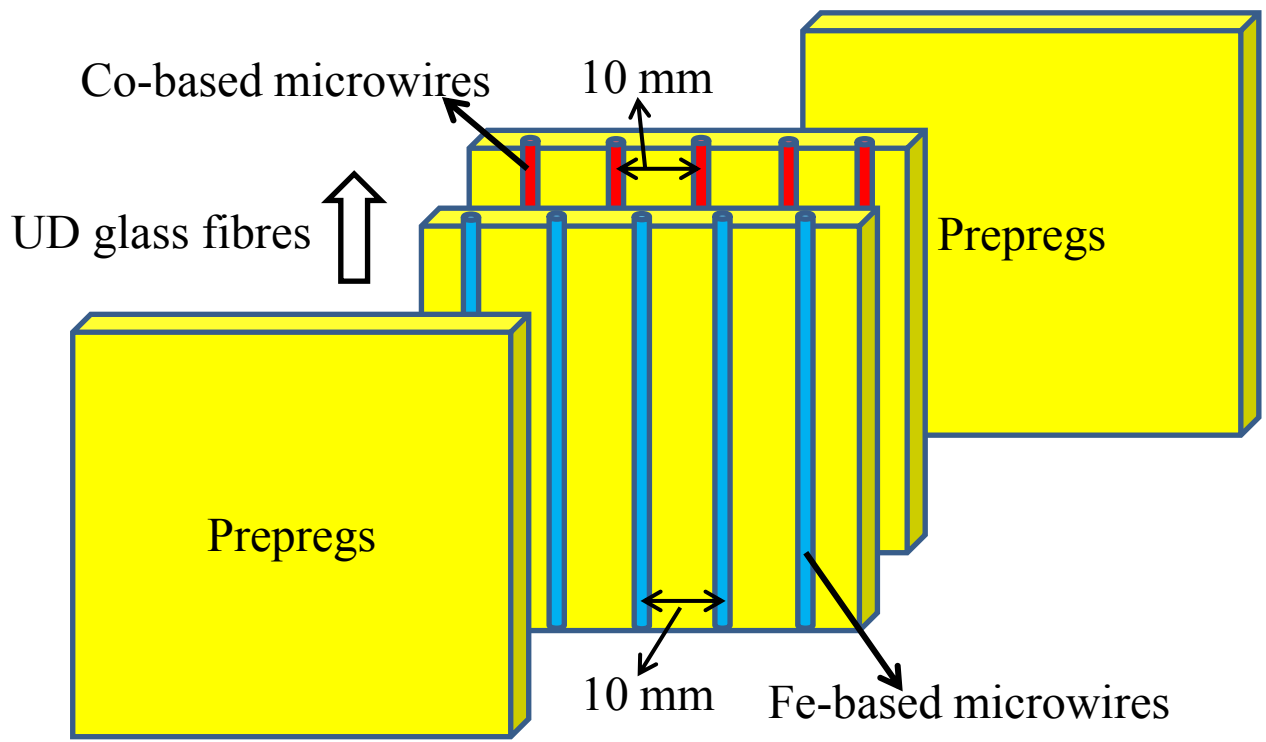

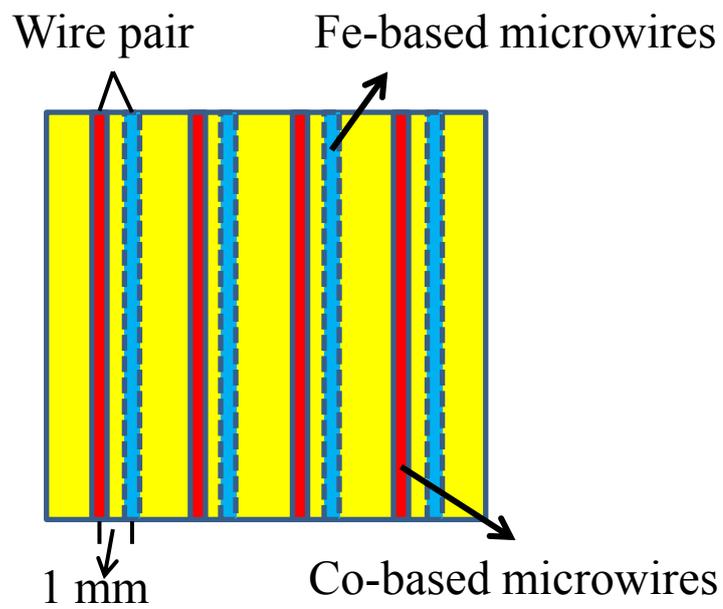

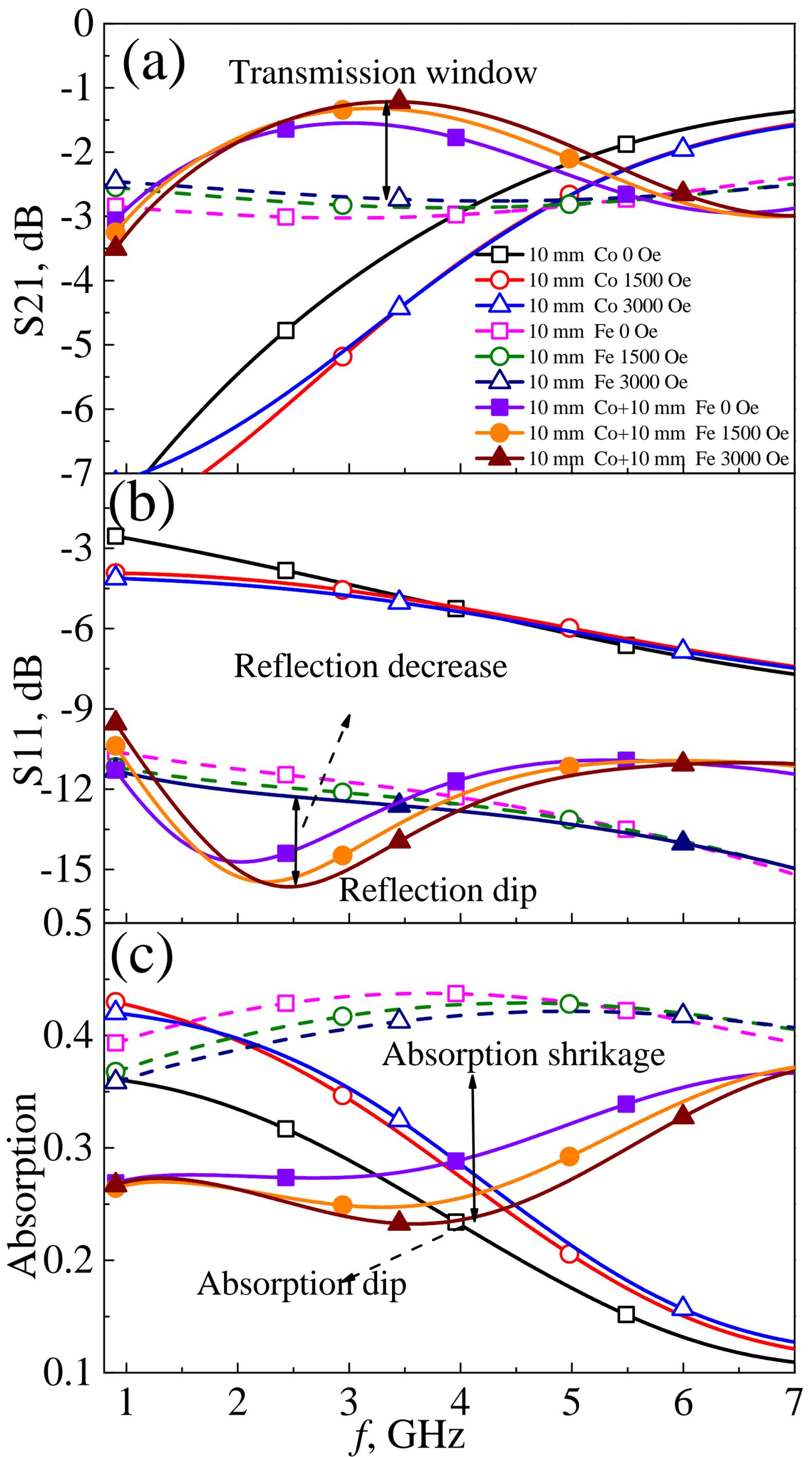

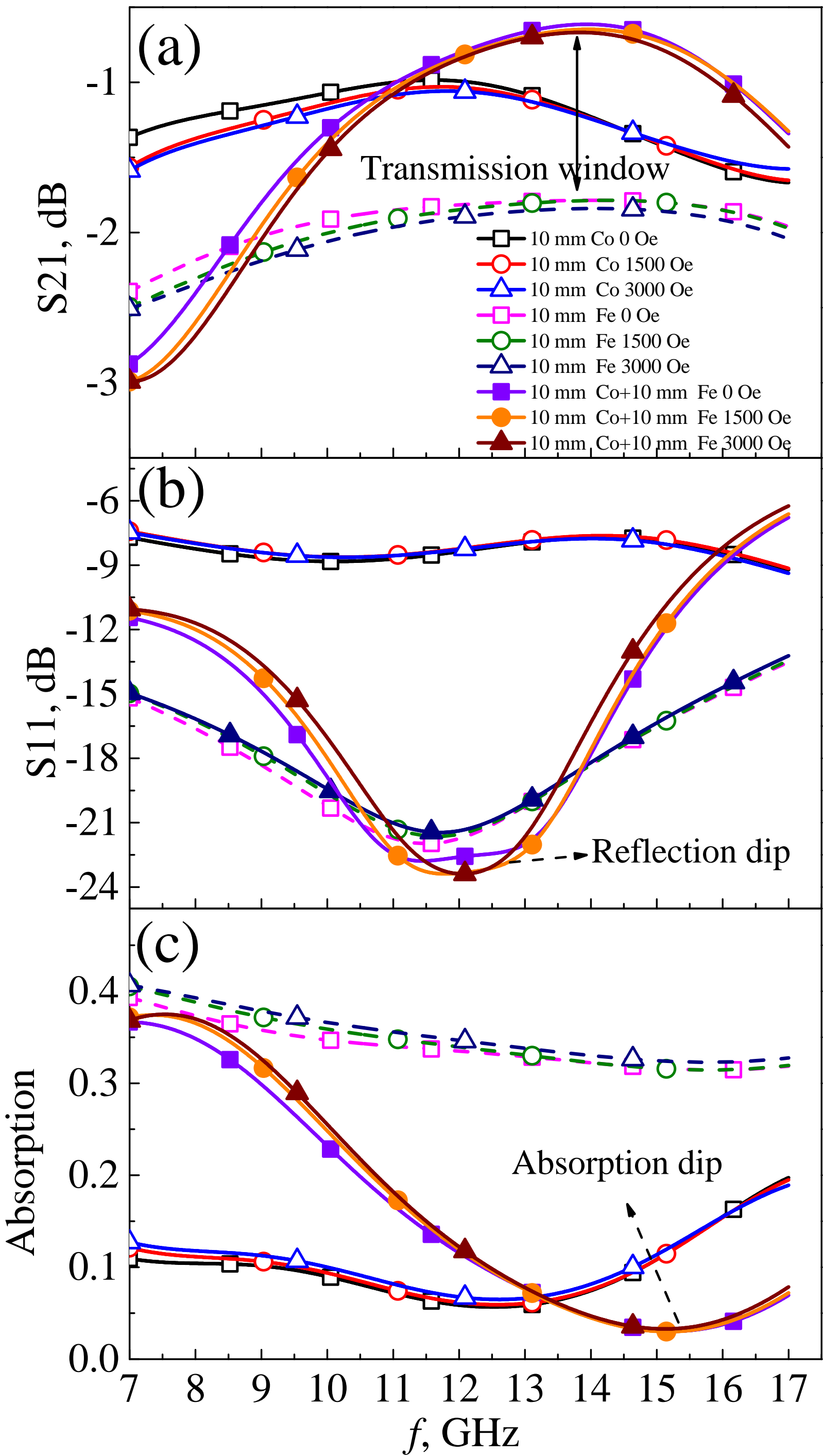

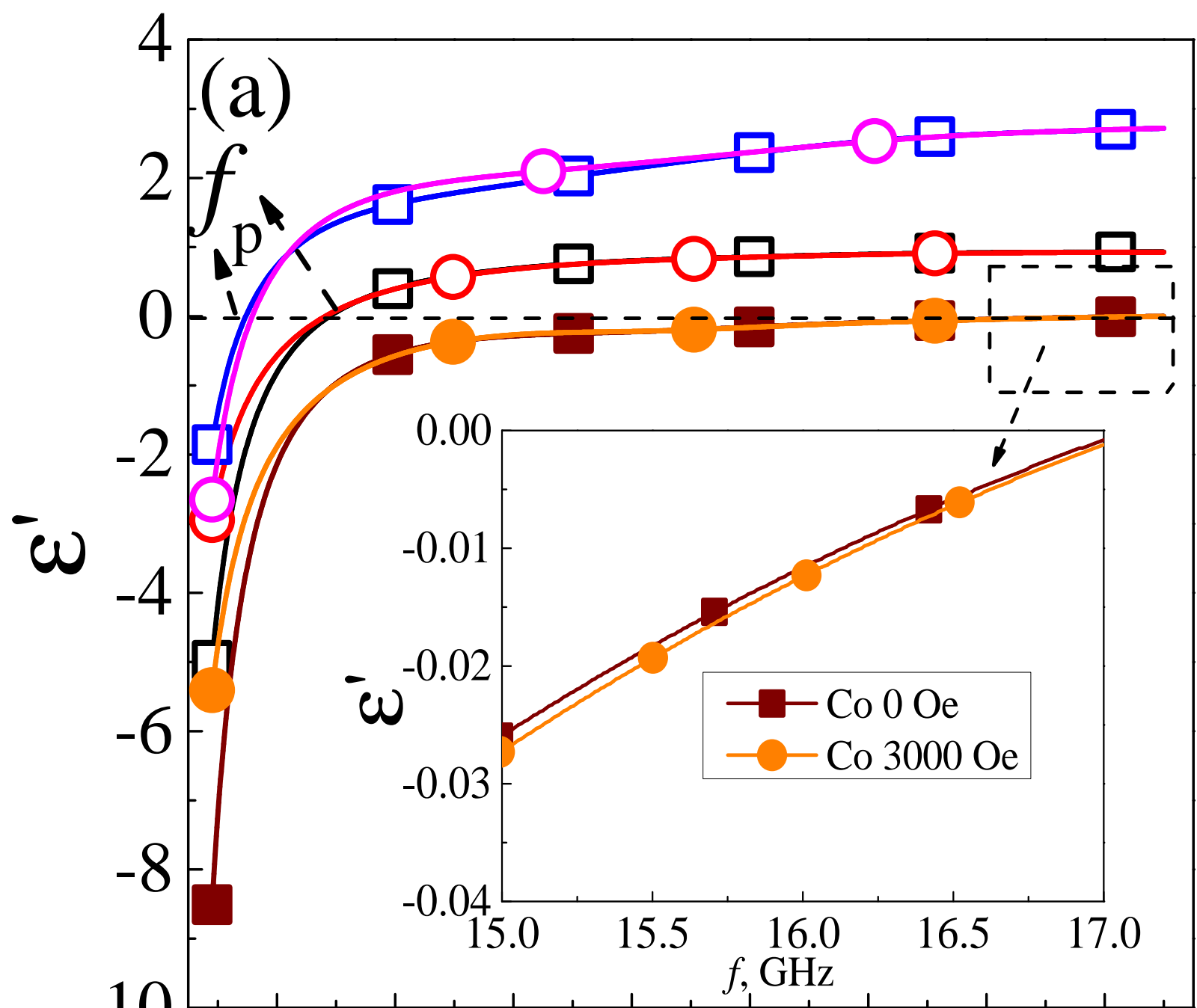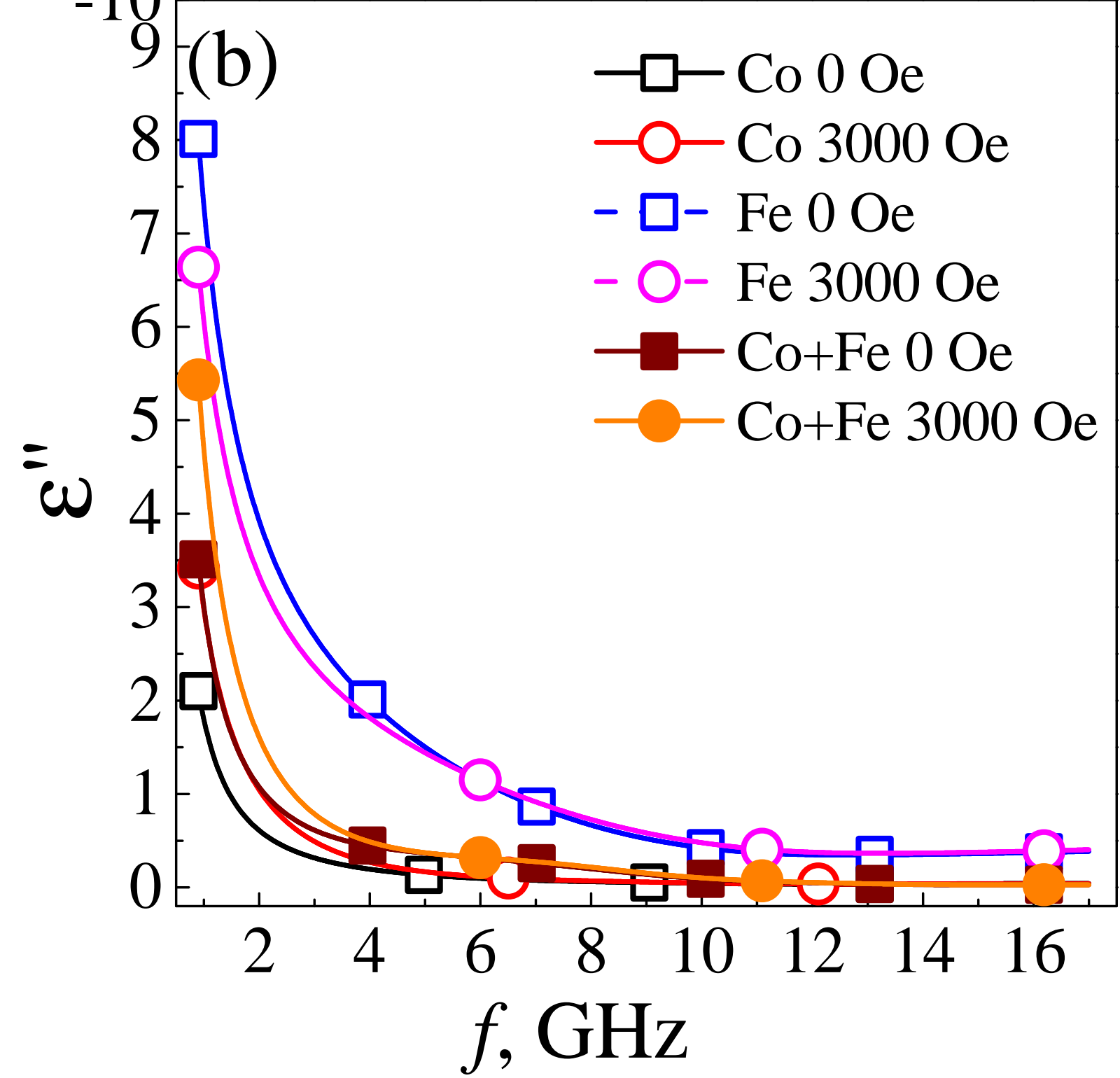